\documentclass{aip-cp}

\usepackage[numbers]{natbib}
\usepackage{rotating}
\usepackage{graphicx}

\begin{document}

% Title portion
\title{Perspectives With The GCT End-to-end
Prototype Of The Small-Sized Telescope
Proposed For The Cherenkov Telescope Array}

\author[aff1]{H. Costantini\corref{cor1}}
\author[aff2]{J.-L. Dournaux}
\author[aff1]{J.-P. Ernenwein}
\author[aff2]{P. Laporte}
\author[aff2]{H. Sol}
\author{for the GCT Team}
\author[aff3]{the CTA Consortium} 

\affil[aff1]{Aix Marseille Univ, CNRS/IN2P3, CPPM, Marseille, France}
\affil[aff2]{Labs LUTH \& GEPI, Observatoire de Paris, CNRS, PSL Research University, Univ Paris Diderot}
\affil[aff3]{See www.cta-observatory.org for full author \& affiliation list}
\corresp[cor1]{Corresponding author: costant@cppm.in2p3.fr}

\maketitle

\begin{abstract}
In the framework of the Cherenkov Telescope Array (CTA), the GCT (Gamma-ray Cherenkov Telescope) team is building a dual-mirror
telescope as one of the proposed prototypes for the CTA small size class of telescopes. The telescope is based on a Schwarzschild-
Couder (SC) optical design, an innovative solution for ground-based Cherenkov astronomy, which allows a compact telescope structure, a
lightweight large Field of View (FoV) camera and enables good angular resolution across the entire FoV. We review the different
mechanical and optical components of the telescope. In order to characterise them, the Paris prototype will be operated during several
weeks in 2016. In this framework, an estimate of the expected performance of this prototype has been made, based on Monte Carlo
simulations. In particular the observability of the Crab Nebula in the context of high Night Sky Background (NSB) is presented.
\end{abstract}

\section{INTRODUCTION}
The Cherenkov Telescope Array (CTA) \cite{cta} is the next major international instrument in very-high energy
astrophysics (20 GeV-300 TeV), providing a sensitivity of one order of magnitude better than
present detectors in this field, as well as significant improvements in angular resolution. The CTA array will consists of 3 different sizes of Imaging Air Cherenkov Telescopes (IACT): 23 m, 12m and 4m diameter dishes and will be located in both the Northern and the Southern hemispheres to observe the whole sky.
The CTA Southern array will include 70 small-sized telescopes (SSTs) spread over an area of few km$^2$ that will dominate the CTA sensitivity in the photon energy range from a few TeV to over 100 TeV. 
The GCT (Gamma Cherenkov Telescope) is one of the proposed telescope design for the SSTs array.  It is based on  a Schwarzschild-
Couder (SC) optical design and couples a large FoV (8 $^{\circ}$) with optimal optical performance \cite{tibaldo}. The first prototype of the GCT camera is based on Multi Anode Photomultipliers (MAPMTs). The GCT prototype has been installed at the Meudon Observatory close to Paris and has been operated at the Meudon site  in November 2015. For the first time Cherenkov light from cosmic showers has been detected by a CTA telescope prototype \cite{jason}. A new observation campaign is foreseen at the end of 2016.

In the first part of this contribution a review of the mechanical and optical components of the telescopes will be given. In a second part the Monte Carlo simulation of the Meudon prototype will be presented together with the expected performance and the observability of the Crab Nebula in the context of high NSB.

\section{THE MECHANICAL STRUCTURE}

\begin{figure} [!h]
  \centering
  \includegraphics[width=0.45\linewidth]{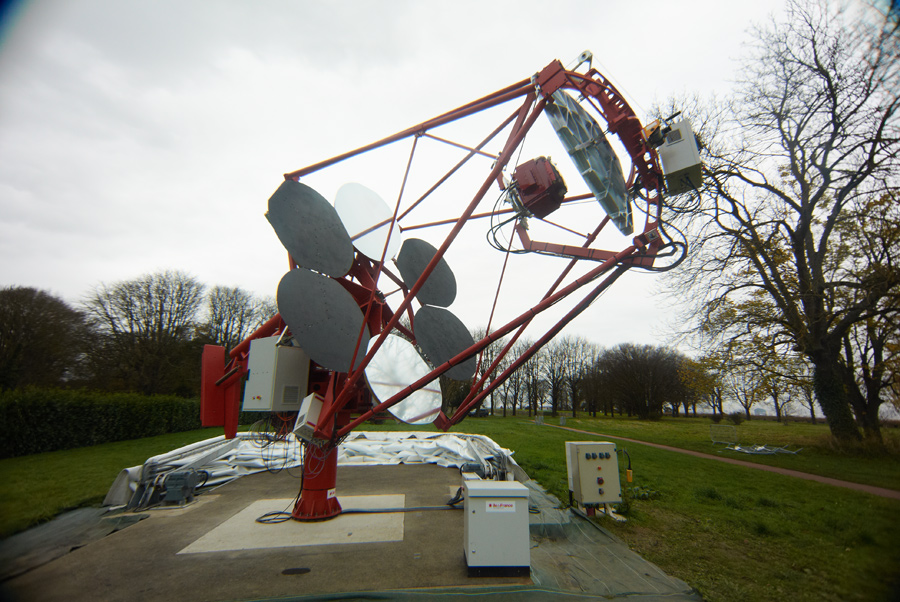}
  \caption{The GCT prototype equipped with its Cherenkov camera on the Meudon campus of the Observatoire de Paris in November 2015. }
  \label{fig:telescope}
\end{figure}

\begin{figure} [!h]
  \includegraphics[ width=0.5\linewidth]{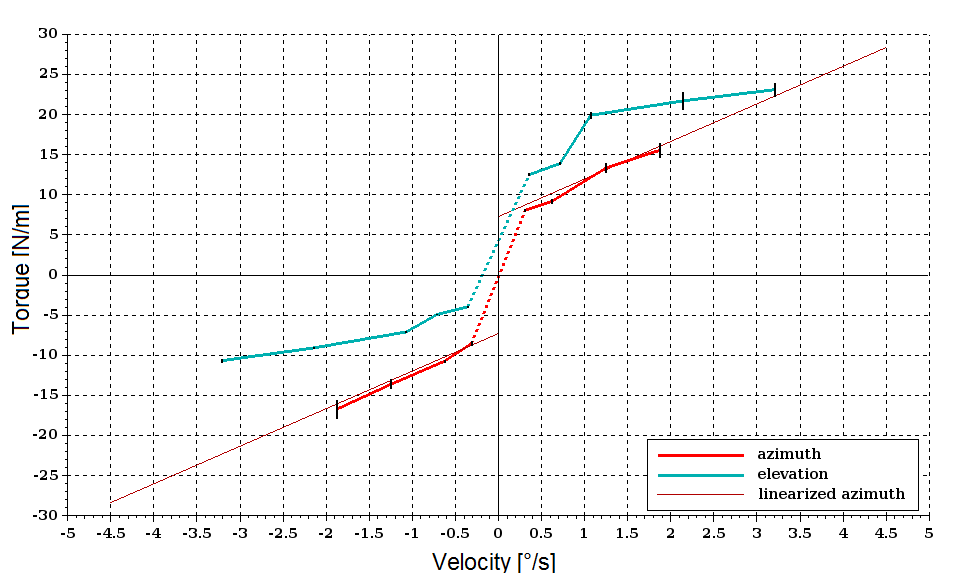}
   \includegraphics[ width=0.5\linewidth]{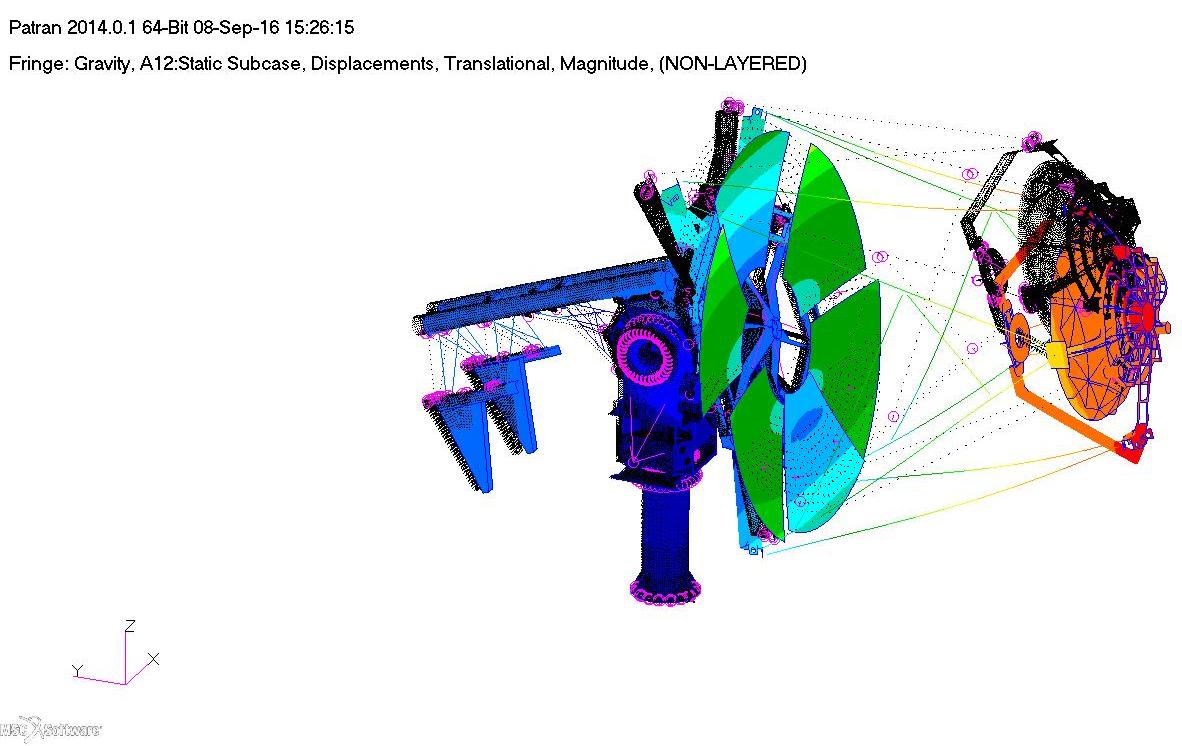}
    \includegraphics[ width=0.06\linewidth]{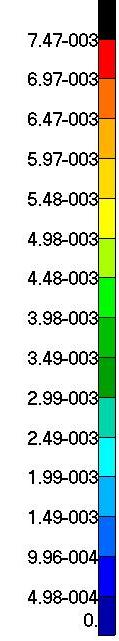} 
       
          \caption{Required torque versus axis velocity (left). Plot of gravity-induced displacements (in magnitude and in meters) for 20$^{\circ}$ elevation (right). The non-deformed structure appears in dotted lines.The distortions on the deformed structure are magnified to be more visible.}
  \label{fig:torque}
\end{figure}

The mechanical structure of the GCT has been designed with the aim of developing a lightweight structure and easing the production, assembly and maintenance phases. It is composed of (i) the telescope base (tower), (ii) the Alt-Azimuth Sub-system (AAS), (iii) the Optical Support Structure (OSS), (iv) the camera access and (v) the foundations. The mechanical design studies started in 2011 at the Observatoire de Paris in France and ended in 2014. A recent detailed description of this structure can be found in \cite{spiemecha16}.
The AAS provides the azimuth and elevation motions of the telescope ($\pm$270$^{\circ}$ and from -5$^{\circ}$ to 90$^{\circ}$ respectively) and includes two similar drive systems for azimuth and elevation. The use of identical drive systems helps to decrease the cost of the purchase, maintenance and spares management. An absolute angle encoder is included in the drive systems in order to give access to the orientation of the telescope, which can be known with an accuracy of $\pm$  2 arcsec.
The camera access includes an innovative camera removal mechanism, which eases the access to the camera for maintenance operations thanks to an ingenious procedure consisting of unlocking the upper arm when the telescope points to the horizon, and rotating down the camera structure. In the removal position, the focal plane is at 1 m from the ground and its access becomes easier.
The majority of the structural frame is currently made of standard carbon steel. Some items are in aluminium grade 6 or in stainless steel. The mass of the telescope (including mirrors and camera) is 8.1 tons. The telescope is shown in Fig.~\ref{fig:telescope}.
The manufacturing of the components of the assemblies was subcontracted between 2012 and 2015. The integration of the AAS with a perpendicular setting system required two weeks. The assembly of the mechanical structure from pre-assembled manufactured assemblies was accomplished in two days while the integration of the camera on the telescope took only 15~minutes thanks to the camera removal mechanism \cite{spiemecha16}.
Tests and analyses have been performed to verify that the telescope structure fulfils the specifications defined by CTA \cite{spiemecha16}. The most constraining tests concern the movement of the telescope. Figure \ref{fig:torque} left shows the first measured data. The slight unbalance of the elevation is related to late developments in the mechanical structure and shall easily be solved with a tuning of the counterweight. The linearised azimuth plot shows that at the maximum velocity of 4.5$^{\circ}$/s, the required torque is lower than 30 Nm, leaving a good margin compared to the continuous torque of 51 Nm provided by the current motor.
Finite-element analyses have been performed to check that the telescope structure meets the requirements in environment conditions and to demonstrate that GCT meets CTA specifications in the operational state. Figure~\ref{fig:torque} right plots the gravity induced displacements.
%\begin{figure} [!h]
  %\centering
  %\includegraphics[ width = \columnwidth]{fea.jpg}
  %\caption{Plot of gravity-induced displacements (in magnitude and in meters) for 20 $^{\circ}$ elevation. The non-deformed structure appears in dotted lines.}
  %\label{fig:fea}
%\end{figure}

\section{MIRRORS}
\begin{figure} [!h]
  \centering
  \includegraphics[ width=0.2\linewidth]{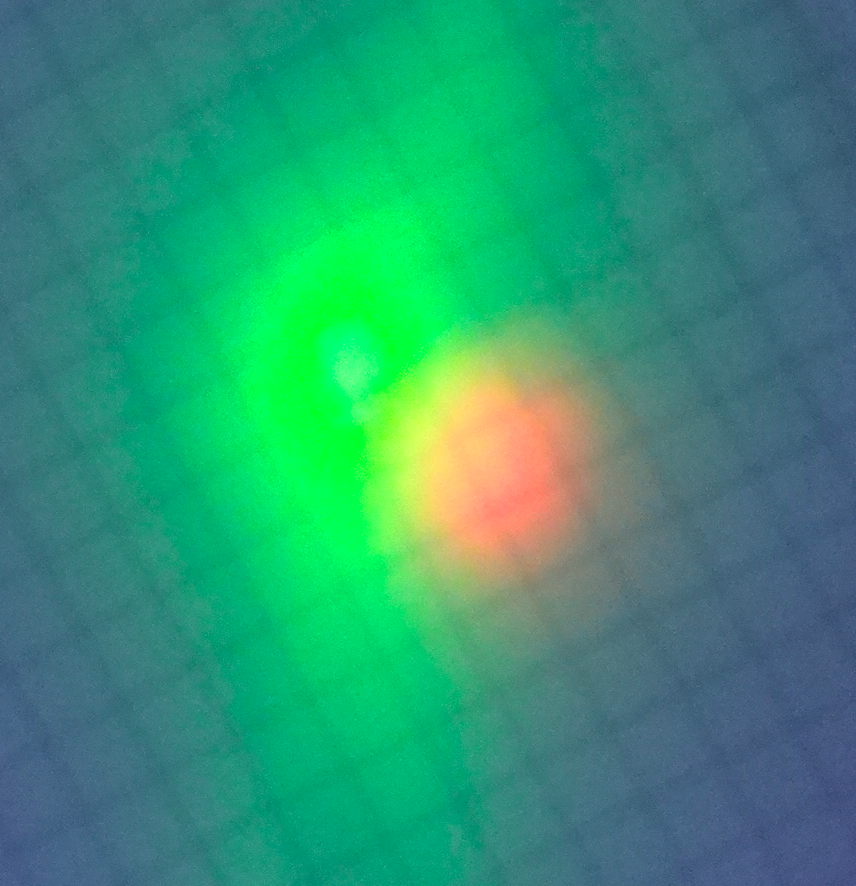}
  \caption{Discrepancy between mechanical (red spot) and optical (green spot) axes. A square sizes 5 mm.} 
  \label{fig:spot}
\end{figure}

The GCT optical design is based on a dual-mirror Schwarzschild-Couder (SC) configuration and has been made in order to have a very small point spread function (PSF) over the whole FoV of 8 $^{\circ}$. The two mirrors, M1 and M2, are strongly aspheric, with a diameter of 4 and 2 meters and closest curvature radius of 9.7 and 2.1 meters respectively. The M1 mirror is tessellated into six identical panels. The GCT team has preferred a non-conventional solution based on machined aluminium lightweight mirrors to develop the prototype mirrors. This solution offers the advantages of an easy and reliable way to obtain the required profile shape and of a light maintenance compared to glass mirrors often used in VHE astronomy.
Because of the absence of actuators on the prototype, the alignment of the optical components of the telescope cannot be optimised. Nevertheless, the precision of the optical measurements made with the telescope was determined by comparing spots induced on the focal plane by two laser beams materialising optical and mechanical axes \cite{spie16opt}. Figure \ref{fig:spot} shows the discrepancy between the two spots and defines a gap of about 7 mm corresponding to an offset of only 10 arc-minutes between these axes.

\section{MONTE CARLO SIMULATION OF THE GCT PROTOTYPE}

A Monte Carlo simulation of the GCT prototype has been set up. The main goal of this simulation is to compare Monte Carlo data to real data such as width and length of the images formed by the atmospheric showers on the camera and compare cosmic rays counting rate with simulated  one. The purpose of these comparisons is to prove the good understanding of the GCT prototype characteristics such as the detection efficiency,  the optical point spread function, the mirrors reflectivity, the photodetector and readout characteristics. This study is important since the telescope characteristics are used in global CTA Monte Carlo simulations to predict CTA performance.
Once basic features are understood a science verification test would be the observation of well known very bright $\gamma$-ray sources such as the Crab nebula. In this framework  the observability of the Crab Nebula  in the context of very high NSB at  the Meudon sky has been evaluated.

The GCT prototype has been simulated considering:
\begin{itemize} 
\item 6 circular petals for the primary mirror (in the current status of the GCT prototype only 2 petals are mounted on the structure) 
\item 2-mirrors reflectivity $\sim$ 50\% 
\item optical point spread function = 5 mm (80\% containment diameter)
\end{itemize}

The last two items have been been obtained from preliminary measurements on the prototype. When more  precise values will be available from future measurements on site the analysis will be updated. 

A set of 4$\times$10$^5$ gammas in the energy range 500 GeV-100 TeV and 5$\times$10$^7$ protons in the energy range 315 GeV-300TeV have been simulated using the Corsika simulation code \cite{corsika} following the shower development until the Meudon site at 162 m above sea level. The NSB has been considered at the level of 200 MHz which has been measured at Meudon in case of clear and moonless nights using an Unihedron SQMLE \cite{sqm}. 
$\gamma$-rays have been generated as originating from a point source while protons have been produced as a diffuse flux in a cone of 20$^{\circ}$  full aperture which is much larger than the telescope FoV (about 8$^{\circ}$). 
Considering the Crab elevation at the Meudon site and the number of moonless nights, it has been evaluated that the Crab stays at a Zenith $<$ 33 $^{\circ}$ for about 255 hours in a year. The prototype has been therefore considered at a Zentih of 30$^{\circ}$ for this analysis.
The tracking of the Cherenkov photons at the telescope level until the detection at the camera including the electronics simulation and the telescope trigger has been done using the \textsl{simtel\_array} \cite{simtel} package. The GCT camera is able to record the full waveform of the signal \cite{jason, tibaldo, target}. Thanks to this feature the charge is computed performing the integration of the waveform around the local maximum of the signal. By considering a small window around the local maximum the contribution of NSB is drastically reduced.
 The  simulated  data  have  been  analysed  with  the \textsl{read\_cta} package \cite{konrad}. Shower images have been cleaned
with a standard two fixed levels cleaning algorithm. The chosen cleaning level was 10 photoelectrons for the higher threshold and 5 photoelectrons for the lower one. 

\subsection{Mono Reconstruction}

\begin{figure}    
\includegraphics[width=0.4\linewidth]{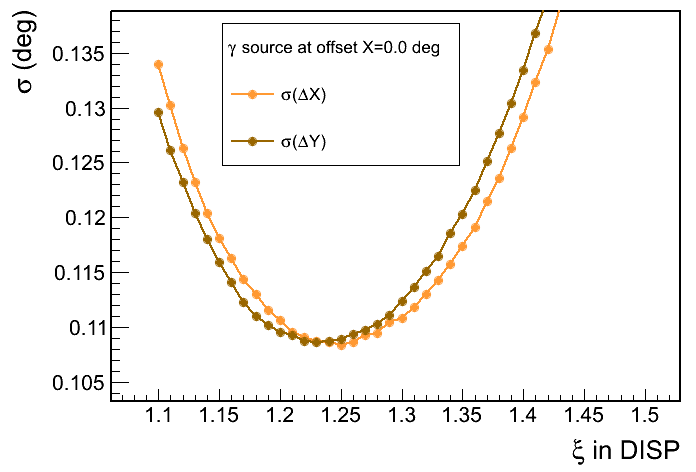}
\includegraphics[width=0.4\linewidth]{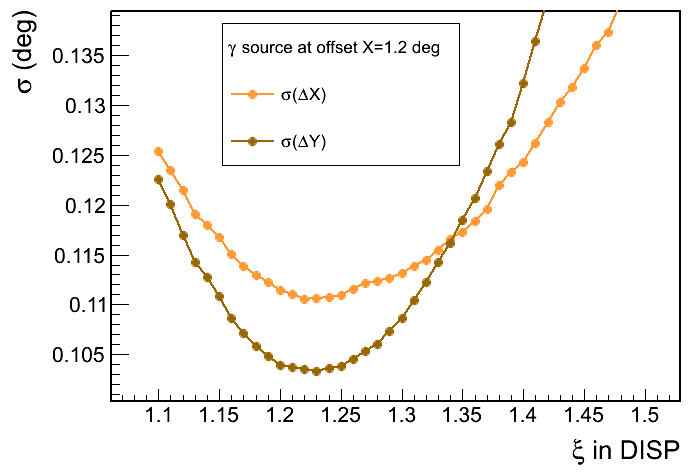}
\caption{ x-y distribution spread from a gaussian fit ($\sigma$) as a function of the $\xi$ scaling parameter  in the case of $\gamma$-ray source at the center of the camera (left) and  at an offset of 1.2 $^{\circ}$ with respect to the camera center (right).}
\label{fig:csi1}
\end{figure}

The location of the $\gamma$-ray source on the focal plane has been determined using the method described in \cite{whipple,bigongiari}.The position of the source lies on the major axis of the ellipse in the direction indicated by the asymmetry
of  the  image which is obtained by the skewness along the axis.  The  distance  of  the  source  from the
image center of gravity, the so-called \textsl{disp}, and the elongation of
the image, defined as the ratio of the image width and length, depend upon the impact parameter of the shower. In a first approximation one can assume a linear relationship between the \textsl{disp} parameter and the elongation: $disp = \xi(1-width/length)$, where $\xi$ is a scaling parameter to be determined with MC simulations by searching the minimal spread of the x-y distribution of the reconstructed source position points. For this purpose several $\gamma$-ray source positions have been used in \cite{bigongiari}. In the present work the optimal $\xi$ is searched at offsets of 0 $^{\circ}$ and 1.2 $^{\circ}$. As shown in Figure \ref{fig:csi1}  a value of $\xi$=1.23 is found for both offsets. This value is also found in \cite{bigongiari} which considered a similar analysis for the ASTRI prototype \cite{ASTRI-gamma2016}.

%\begin{figure}
%\includegraphics[width=0.4\linewidth]{xi_optim_12deg.png}
%\includegraphics[width=0.3\linewidth]{psf_12.png}
%\caption{x-y distribution spread from a gaussian fit($\sigma$) as a function of the $\xi$ scaling parameter  and point spread function in the case of $\gamma$-rays source at an offset of 1.2 $^{\circ}$ with respect to the camera center .}
%\label{fig:csi2}
%\end{figure}

\subsection{Gamma-hadron separation}
For gamma-hadron separation a Multi Layer Perceptron  (MLP) Neural Network has been used with 4 inputs, 35 hidden neurons and 1 output. This software has been developed by one of the authors. It has been used and evaluated by R.K. Bock and collaborators \cite{gaug} who compared different  multidimensional event classification methods in the framework of IACTs. 

\begin{figure}

  \includegraphics[width=0.25\linewidth]{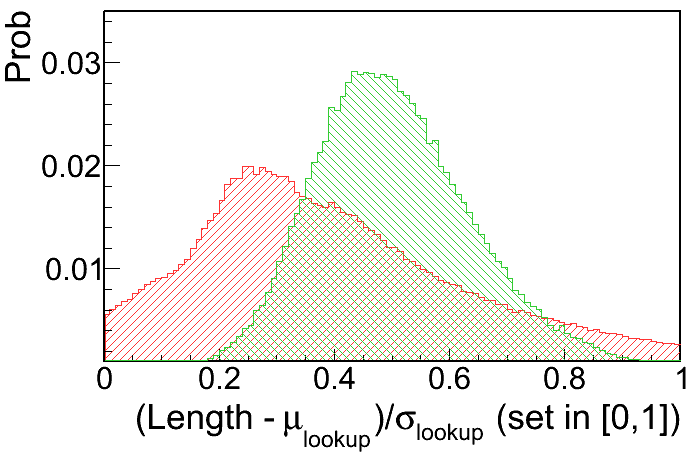}  %\end{minipage}
%\begin{minipage}{.5\textwidth}
 \includegraphics[width=0.25\linewidth]{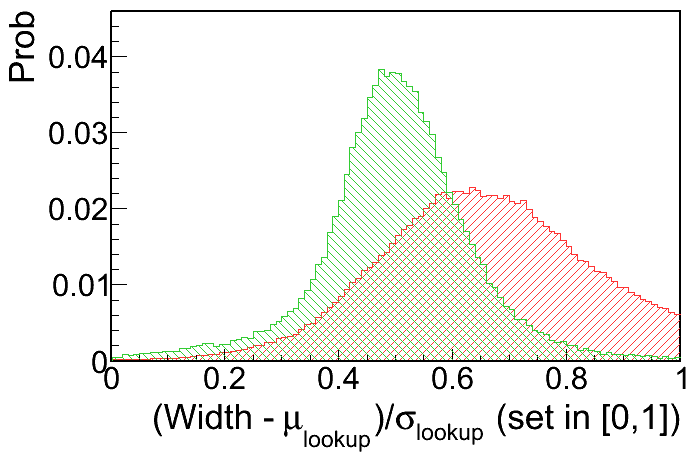}
   \includegraphics[width=0.25\linewidth]{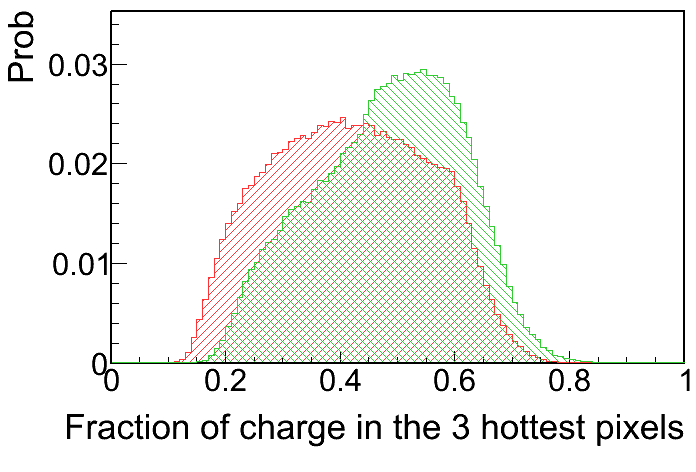}
    \includegraphics[width=0.25\linewidth]{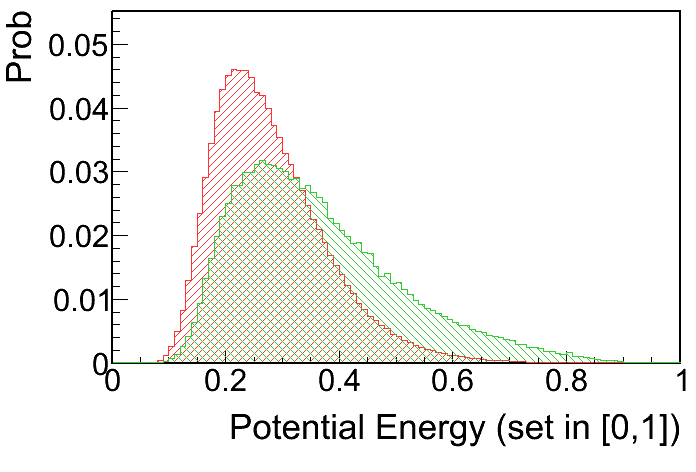}
        \label{fig:inputs}
%\end{minipage}
\caption{Input variables for the MLP Neural Network}
\end{figure}

Two of the four input variables are the scaled width and length of the camera image defined as $(width-\mu_{width-lookup})/\sigma_{width-lookup}$ and $(length-\mu_{length-lookup})/\sigma_{length-lookup}$ respectively.
 $\mu_{x-lookup}$ and $\sigma_{x-lookup}$ are the mean and the RMS of the distribution stored in a lookup table with image charge and dist (distance between the camera center and the center of gravity of the image) as inputs.
The other two inputs are topological variables reflecting mainly the concentration of the illuminated pixels in the camera. One is defined  as the ratio between the charge of the 3 hottest pixels and the total charge of the image and the other is defined  as $\frac{1}{N}\sum_{N_{pairs}}Q_{i}Q_{j}/r_{ij}$ ( a kind of electrostatic energy of the charge distribution). In Figure \ref{fig:inputs} the distributions of the four variables are shown for both $\gamma$-rays (green distribution) and protons (red distribution). The distributions have been rescaled into [0,1] to be used as inputs for MLP Neural Network.

\begin{figure}
 
  \includegraphics[width=0.4\linewidth]{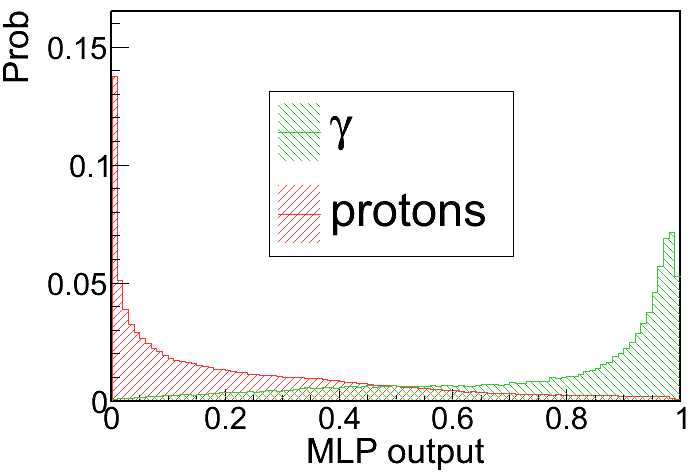}
  %\end{minipage}
%\begin{minipage}{.5\textwidth}
% \includegraphics[width=0.4\linewidth]{rej_eff.png}
        \label{fig:MLP}
%\end{minipage}
\caption{MLP output variable .}
\end{figure}

The MLP output variable is shown in Figure \ref{fig:MLP} .
For the $\gamma$-ray source search a cut on the MLP variable and on the $\theta^2$ (squared distance between the shower reconstructed position from the Mono Reconstruction and the pointed source, here at the center of the camera) has been applied.

\subsection{The Crab nebula observation significance: single OFF method}
To determine the number of $\gamma$-ray events detected from the Crab nebula the differential flux from \cite{holder} has been adopted:
dN/dE=1.79$\times$10$^{-10}$( E / 0.521 TeV)$^{-2.10-0.24\cdot ln(E/0.521 TeV)}$TeV$^{-1}$cm$^{-2}$s$^{-1}$.
For the expected proton background events the following differential flux has been considered \cite{pdg}: dN/dE=1.13$\times$10$^{-5}$cm$^{-2}$ s$^{-1}$sr$^{-1}$TeV$^{-1}$(E/1TeV)$^{-2.7} $.

\begin{figure}
 \includegraphics[width=0.35\linewidth]{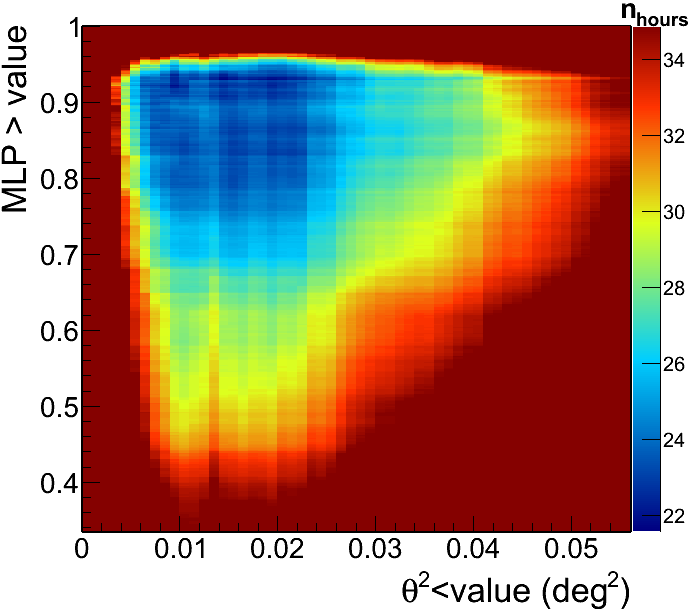}
 \includegraphics[width=0.4\linewidth]{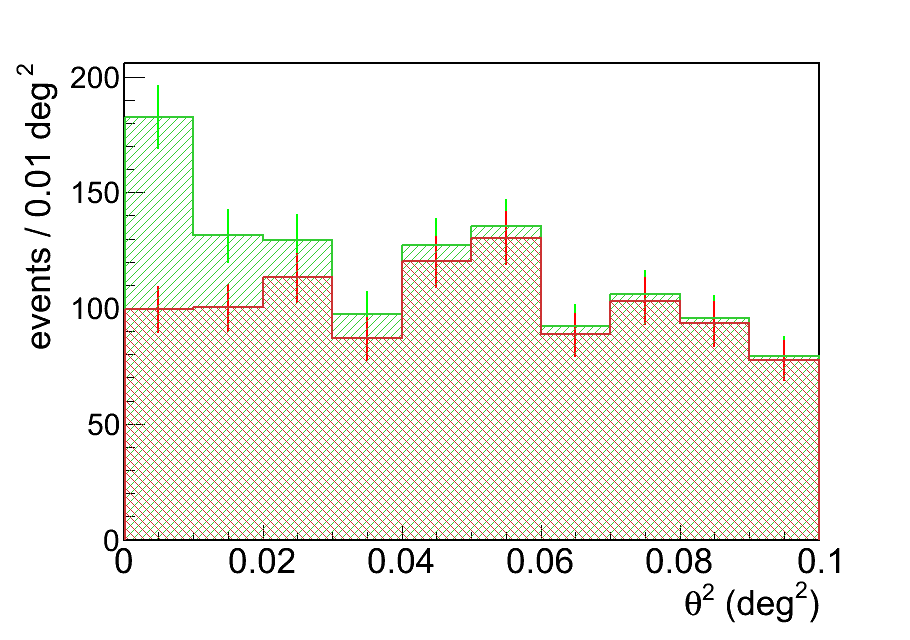}
\label{fig:singleOFF}
\caption{Time required for a 5$\sigma$ detection of the Crab nebula as a function of the MLP output variable and $\theta^{2}$ (left). $\theta^{2}$ distribution for gammas and protons after MLP cut in 24 hours observation (right). }
\end{figure}

The aim of this work has been to determine the time needed to observe the Crab nebula with a 5 $\sigma$ significance according to Li-Ma \cite{lima}. To this end two methods have been adopted. In both cases the Crab is observed on axis. In a first rather conservative method the background has been estimated considering an OFF-source observation of the same duration as the ON-source observation. The cuts on the MLP output parameter and $\theta^{2}$ have been chosen to minimise the observation time for a 5 $\sigma$ significance observation of the Crab nebula (see Figure \ref{fig:singleOFF} left). The obtained cuts are : $\theta^{2}<0.02$ deg$^2$ and MLP$>0.9$. The corresponding $\gamma$-ray efficiency is 0.54 while the proton rejection is 0.98.
The angular resolution of the analysis, defined as the radius containing 68\% of the events, is 0.2 $^{\circ}$.
 The required time for a 5$\sigma$ observation is 24 hours of ON-source observation and 24 hours of OFF-source observation.
 The $\theta^{ 2} $ distribution of the detected gammas and protons  in 24 hours is presented in Figure \ref{fig:singleOFF} right. After the final cuts N$_{\gamma}$=114 and N$_{p}$=200. The Signal over Background ratio S/B is 0.6.

\subsection{The Crab nebula observation significance: ring background method}
A ring background method has been adopted to reduce the time needed for a 5 $\sigma$ observation of the Crab. It consists in defining a ring around the central ON zone ($\theta^{2} <$0.02 deg$^{2}$) to evaluate the proton background in the ON zone. The inner radius of the ring should be large enough to avoid $\gamma$ contamination: $\theta^2 >$0.1 deg$^2$. Taking into account the $\gamma$-like proton acceptance decrease across the FoV, the outer radius is chosen to cumulate an effective collection area for $\gamma$-like protons 10 times larger than the ON zone: $\theta^2<0.55$ deg$^2$.

%In Figure \ref{fig:ring}  (nous avons pas la place pour ces images , aussi si on mets l'image 1D donc je propose de les supprimer) left the acceptance for $\gamma$-like protons is shown and
%It should be noticed that the acceptance of $\gamma$-like protons depends strongly on the $\theta^2$ variable due essentially to the mono analysis technique used. Therefore to determine the effective collection area used for the OFF-zone determination, one has to  take into account this dependence.
  After $\gamma$-hadron separation cuts a proton rate of 1.4 p/min is obtained in the ring while the $\gamma$ rate is of 0.4 $\gamma$/hour therefore the ring zone can be considered essentially $\gamma$ free.  
After 13 hours of ON-source observation, a 5 $\sigma$ observation significance is reached with N$_{\gamma}$=62, N$_{p}$=1120/10 and a Signal over Background ratio S/B=0.6.
An important remark is that OFF data have to be taken to check the $\gamma$-like proton acceptance obtained from simulation across the FoV. Considering a rate of 1.4 p/min inside the ring, 30 hours of pure OFF data would allow an acceptance measurement at a level of 10\%.

\section{CONCLUSIONS}

The GCT prototype has been built and operated  for the first time at the end of 2015. It is a full prototype that includes the mechanical structure to move the telescope in zenith and azimuth, the optical system and a first camera prototype based on MAPMs. 

A Monte Carlo simulation of the prototype has been performed, its first goal being to compare simulated Cherenkov images  with real data taken by the GCT camera during the next observation campaign at the end of 2016. Considering the full 6-petals mirror prototype, the observability of the Crab nebula has been evaluated. The time needed for a 5 $\sigma$ observation ranges from 13 to 24 (+ 24 OFF zone) hours depending on the method. 

\section{AKNOWLEDGMENTS}
We gratefully acknowledge support from the agencies and organizations under Funding Agencies at \textsl{www.cta-observatory.org}

%\begin{figure}
 %\includegraphics[width=0.35\linewidth]{acceptance_pour_ring_method.png}
 %\includegraphics[width=0.32\linewidth]{annulus_cuts.png}

%\label{fig:ring}
%\caption{Acceptance for $\gamma$-like protons (left). Protons (red dots) and $\gamma$ (green dots) distribution in the ON-zone and ring background zone after quality cuts.}
%\end{figure}

% References

\nocite{*}
\bibliographystyle{aipnum-cp}%
%\bibliography{sample}%

\end{document}